\documentclass[3p,times]{elsarticle}

\usepackage{ecrc}

\volume{00}

\firstpage{1}

\journalname{Nuclear Physics A}

\runauth{G.S. Jackson and W.A. Horowitz}

\jid{npa}

\jnltitlelogo{Nuclear Physics A}

\usepackage{amssymb}
\biboptions{square,comma,numbers,sort&compress}

\usepackage[figuresright]{rotating}

 \def\be{\begin{eqnarray}}	 	\def\ee{\end{eqnarray}}
 \def\bea{\begin{eqnarray}}	 	\def\eea{\end{eqnarray}}
 \def\bean{\begin{eqnarray*}}	\def\eean{\end{eqnarray*}}

 \def\bm#1{\mbox{\boldmath$#1$}}
 \def\d{\textrm{d}}

 \newcommand{\eq}[1]{(\ref{#1})} 
 \def\ie{i.\,e.}	% id est (that is)

\begin{document}

\begin{frontmatter}



\dochead{}
%% e.g. \dochead{17th International Conference on Dynamical Processes in Excited States of Solids}

\title{Predictions for the Spatial Distribution of Gluons in the Initial Nuclear State}


\author[uct]{G. S.~Jackson}
\ead{greg@wam.co.za}
\author[uct]{W. A. Horowitz}
\ead{wa.horowitz@uct.ac.za}

\address[uct]{
  Department of Physics,
  University of Cape Town,
  Private Bag X3,
  Rondebosch 7701,
  South Africa
}

\begin{abstract}
We make predictions for the $t$-differential cross section of exclusive vector meson
production (EVMP) in electron-ion collisions, with the aim of comparing DGLAP
evolution to CGC models. In the current picture for the high-energy nucleus, nonlinear
effects need to be understood in terms of low-$x$ gluon radiation and recombination as
well as how this leads to saturation. EVMP grants experimental access to the edge
region of the highly-boosted nuclear wavefunction, where the saturation scale for CGC
calculations becomes inaccessible to pQCD. On the other hand, DGLAP evolution requires
careful consideration of unitarity effects. The existing $J/\psi$ photoproduction data
in $ep$ collisions provides a baseline for these theoretical calculations. Under
different small-$x$ frameworks we obtain a measurable distinction in both the shape
and normalization of the differential cross section predictions. These considerations
are relevant for heavy ion collisions because the initial state may be further
constrained, thus aiding in quantitative study of the quark-gluon plasma.
\end{abstract}


\end{frontmatter}

% \linenumbers

\section{Introduction}
\label{}

Precise measurements of the strong force responsible for holding the nucleus together
test the basic properties of matter as predicted by QCD physics. The gluonic structure of the
nucleas can be scanned by a quark-antiquark pair produced from a virtual photon. Such
is the case in exclusive vector meson production (EVMP), where a particular final
state of the dipole is produced. Data from HERA is compatible with existing
theoretical models, and has indicated the
existance of a saturation scale \cite{gbw2}. The
nonlinear effects arise from gluon radiation and recombination which becomes
increasingly important at low-$x$. 

What concerns us here, is whether different underlying physical assumptions for the
evolution of gluon densities in nuclei could be used to arrive at testable predictions
with measurable differences. This adds to the case for future electron-ion collider
facilities, as EVMP provides a fertile testing ground for the gluon distribution
\cite{ck}. We shall begin with an analysis of $ep$ collisions, using
this to confirm that the momentum distributions are evolving correctly. Following this, we
present results for nuclear scattering.  Traditional DGLAP evolution will be compared
to novel CGC physics based on the running coupling BK equation \cite{bk3}
(rcBKC).
The DGLAP models feature impact parameter dependence based on work in \cite{kt}. CGC
based calculations include explicit dependence on the number of overlapping nucleons
at a given impact parameter \cite{albacete}.

\section{The nucleon as a baseline}

Consider the electron-proton interaction $e+p \rightarrow e + V + p$, which
proceeds through the exchange of a virtual photon between the proton and electron. It
is standard to focus on the QCD contribution, namely $\gamma^* + p \rightarrow V + p$.
When $V$ is a vector meson, the momentum fraction is given by
\bean
x &=& \frac{Q^2+M_V^2}{Q^2+W^2}.
\eean
Therefore the squared invariant mass $W^2$, of the produced hadronic matter, may be
used as a proxy for the $x$ evolution.
We shall be using data from \cite{zeus2002, h12005} in order to fix the parameters in
our models. The observation that $\d \sigma/ \d t \sim e^{t B_p}$ gives the total
cross section as a proportion of the forward scattering of the differential cross section,
\bea
\sigma_{tot} (W) &=& \frac{1}{B_p} \frac{\d \sigma^{\gamma^* p \rightarrow Vp}}{\d |t|}
\bigg|_{t=0}.
\label{sigtot}
\eea
A formal description of EVMP in the dipole picture may be
found in \cite{kmw}. For our purpose, the elastic
diffractive cross section may be written in terms of the squared amplitude,
\bean
\frac{\d \sigma_{T,L}^{\gamma^* p \rightarrow V p}}{\d |t|} &=&
\frac{1}{16 \pi} \vert \mathcal{A}_{T,L}^{\gamma^* p \rightarrow Vp} (x,Q^2, \bm
\Delta) \vert^2.
\eean
Where the $Q^2$ is the virtuality of the incoming photon, and ${\bm \Delta}$ is the
momentum imparted to the target (\ie \ $t = - \bm \Delta^2$). The subscripts $T$ and
$L$ refer to the transverse and longitudinally polarised photons respectively. By
considering the timescales involved \cite{msm}, it is justified to write the amplitude
as
\bean
\mathcal{A}(x,Q^2, \bm \Delta) &=&
\int \d^2 \bm r \int \frac{\d z}{4 \pi} (\Psi_{V}^* \Psi)(\bm r, Q^2, z)
\frac{\d \sigma_{q\bar{q}}^p}{\d t}.
\eean
Information pertaining to the gluonic interaction is encoded in the dipole term. The
vector-meson photon overlap $(\Psi_V^* \Psi)$ gives the amplitude for the incoming
photon to split into a $q\bar{q}$ pair and then recombine into a vector-meson.
Kinematic choices for $\bm b$ set it to be Fourier conjugate to the momentum transfer
{\bm \Delta}.

We follow previous work \cite{ck}, and assume that the momentum scale of the
ineraction depends on the magnitude of the dipole separation, $\mu^2 (r) =
C/r^2 + \mu_0^2$. There are existing best-fit values for $C$ and $\mu_0$ in the
literature, however we shall
be obtaining our own values with relevance to the $x$ evolution. A more precise
derivation of this scale is left for future work. Identifying the opacity as
\bean
\Omega =  r^2 F(x, r) T (\bm b), 
\ &\mathrm{where}& \ F(x,r) = \frac{\pi^2}{2 N_c} \alpha_s (\mu^2) xg (x,\mu^2),
\eean
assuming that the $\bm b$ dependence of $xg$ can be neglected.
For a thick target, the dipole cross section has the form
\bea
\frac{\d \sigma_{q \bar{q}}^p}{\d^2 \bm b} (\bm b, \bm r, x) &=&
2 \left[ 1 - \exp \left( - \Omega \right) \right],
\label{GM}
\eea
in accordance with the Glauber-Mueller formula, (henceforth abbreviated ``GM''). We shall also study a
linearised version of this expression with a unitary cut-off, viz.
\bea
\frac{\d \sigma_{q \bar{q}}^p}{\d^2 \bm b} (\bm b, \bm r, x)  &=&
\Omega  \theta \left( 2 - \Omega \right)
+ 2 \theta \left( \Omega - 2 \right),
\label{thetaCK}
\eea
which clearly respects the large $r$ behaviour. This we name ``$\theta$-CK''
\cite{ck}. Comparison between \eq{GM} and
\eq{thetaCK} will indicate the importance of the saturation scale.

\begin{figure}
  \centerline{\includegraphics[scale=0.45]{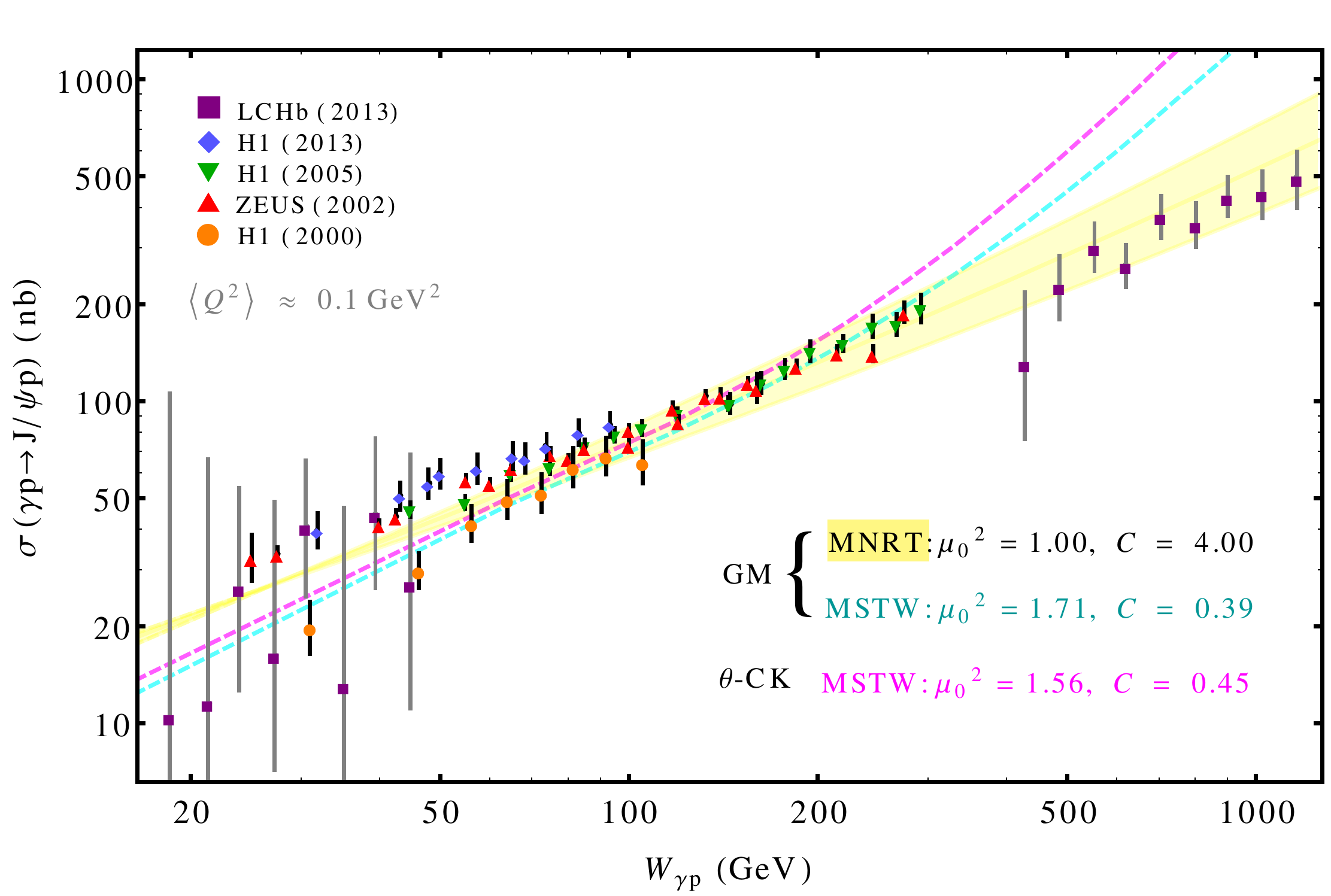}}
  \caption{ (Colour online)
  The theoretical curves show $\sigma (W)$ for a
  mean $Q^2 \approx 0.1$ GeV$^2$, consistent with the \cite{h12005, zeus2002} data. Further
  data points are shown for comparison but have different criterion on the photon virtuality cuts.
}
\label{fig1}
\end{figure}

Our results depicting the total cross section \eq{sigtot} as a function of $W^2$ are
reproduced in Fig.~\ref{fig1}. The parameters $\mu_0$ and $C$ have been left as free,
with best-fit values presented in Fig.~\ref{fig1}. The ``MNRT'' parametrisation is a
simplifaction for the evolutions of the gluon pdfs \cite{mnrt}. Proper DGLAP evolution
is applied using ``MSTW'' code \cite{mstw}. The MSTW produces a run-away result at
large $W$ (small-$x$), due to the fact that the gluon pdfs are unconstrained below $x
\simeq 10^{-5}$.

\section{The nuclear target}

In configuration space, we now treat the nucleus in the transverse plane as a collection
of nucleons with coordinates $\lbrace \bm b_1, \bm b_2, \ldots, \bm b_A \rbrace$. The  DIS event
is characterised by the scattering matrix $S_A(\bm r, \bm b, x)$, which may be
expressed in terms of a product over the nucleons with the independent scattering
approximation.
Positions of the nucleons must be averaged over, in order to talk about
$t$-differential cross sections. The coherent cross section, involves $|\langle
\mathcal{A} \rangle |^2$ and represents the process $\gamma^*+ A_0 \rightarrow V+ A_0$.
The incoherent cross section has the nucleus leaving in an excited state, and is
calculated through the variance $\langle | \mathcal{A} |^2 \rangle - |\langle
\mathcal{A} \rangle |^2$. Presently, our numerics are only capable of resolving the
coherent contribution.
The running coupling BK is caputred through a numerical model laid out in
\cite{albacete}. The nucleon coordinates are generated through a Monte Carlo
simulation.

\begin{figure}
  \centerline{
  \includegraphics[scale=0.45]{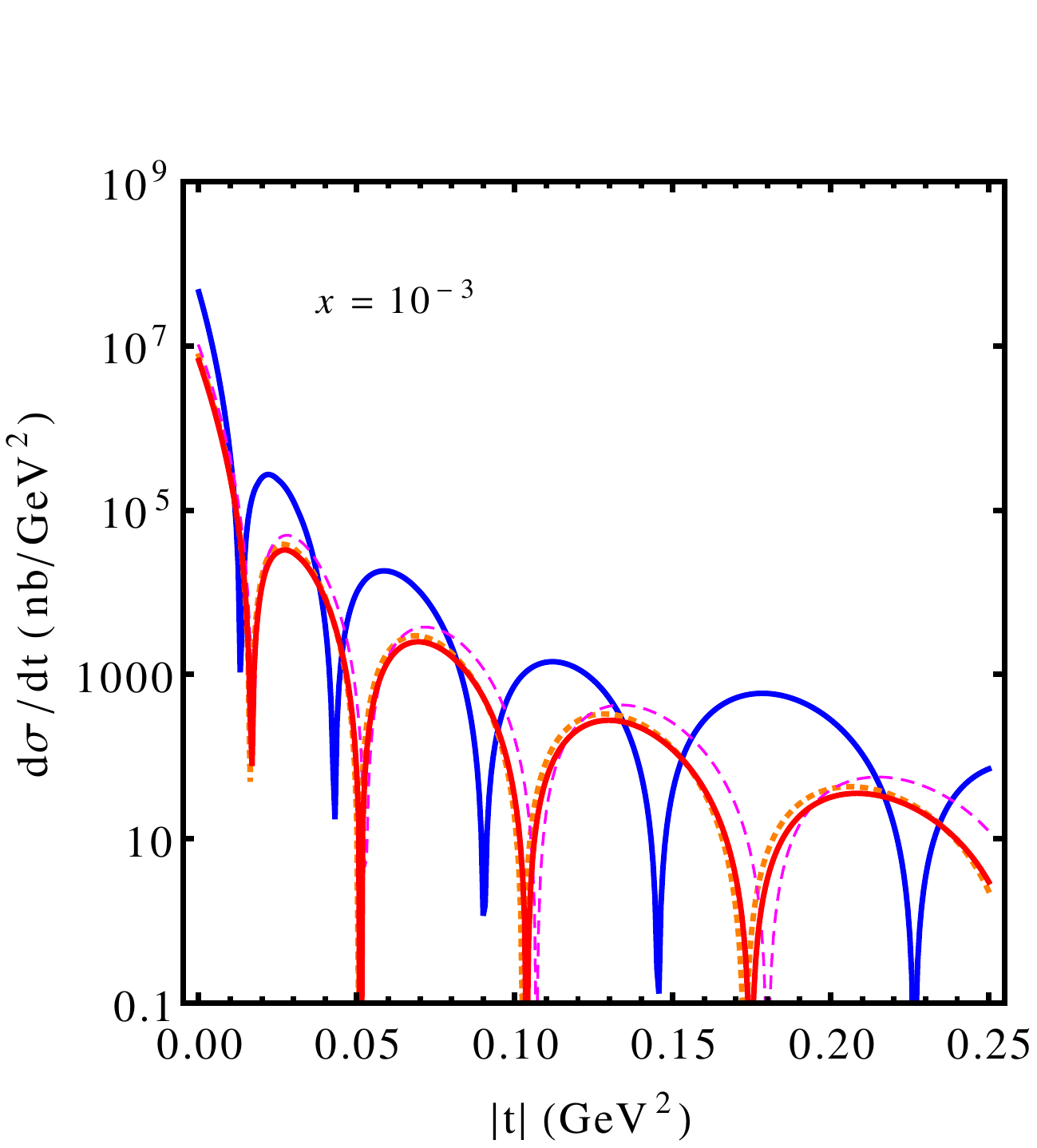}
  \includegraphics[scale=0.45]{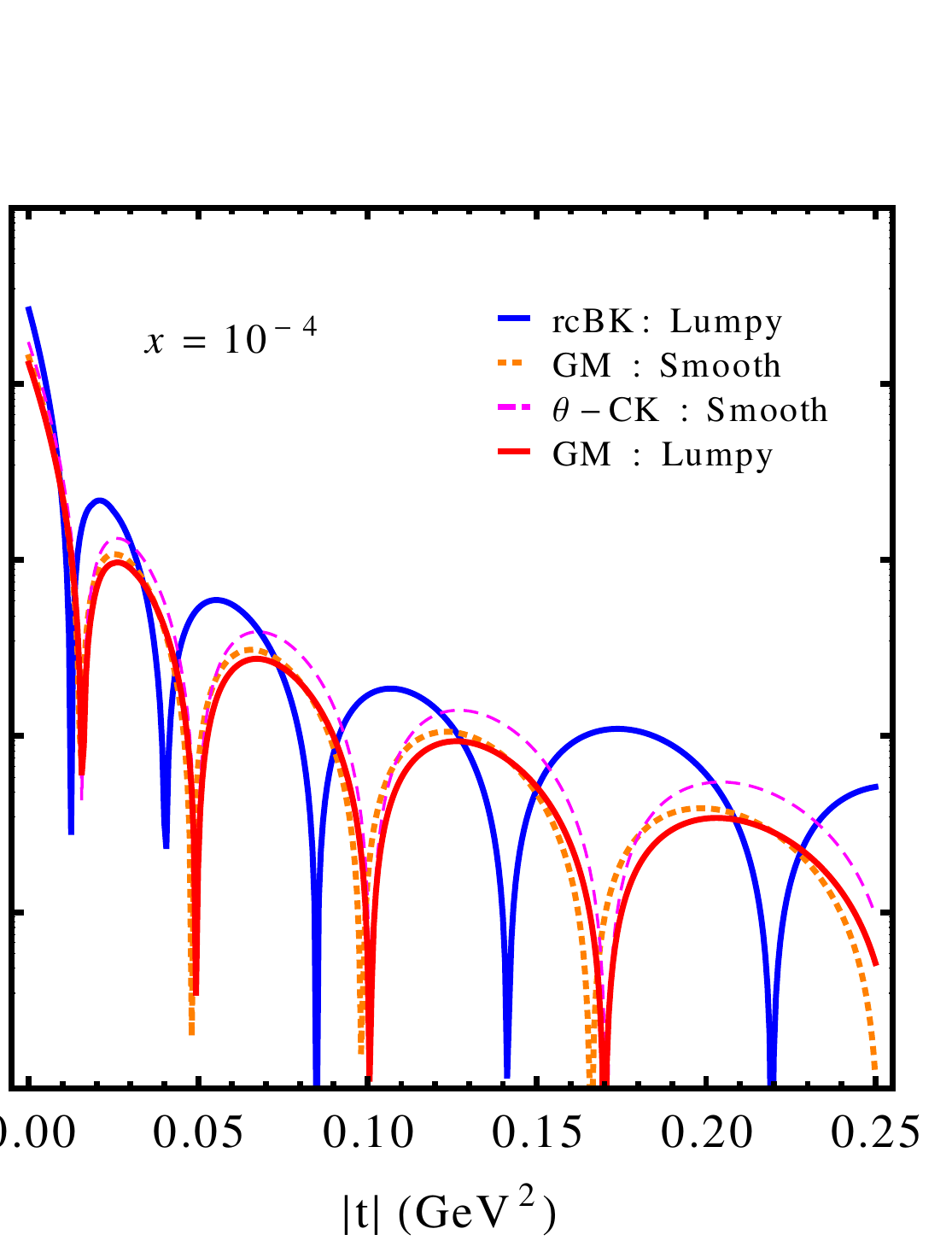}
}
\caption{ (Colour online)
  A comparison between the cross section calculated using the different models
  at (left) $x = 10^{-3}$, (right) $x=10^{-4}$. }
\label{fig2}
\end{figure}

A leading order calculation, under the assumption that the scattering matrix
factorises in combination with the large $A$ limit, justifies the modification of
Eq.\eq{GM} to simply treat the thickness function $T({\bm b})$ as a normalised
transverse Woods-Saxon distribution, denoted by $T_A$. Thus, in this way, the nucleus also inherits
model \eq{thetaCK}. Both of these models average over nucleon coordinates in a
simplified manner and thus describe a ``smooth nucleus''.
In addition, we shall approximate the ``lumpy nucleus'' in the following way.
Averaging the impact parameter cross section involves a convolution of $A$  Woods-Saxon
distributions, by modifying the opacity according to $T(\bm b ) \rightarrow
\sum_{i=1}^A T_p (\bm b - \bm b_i)$.
\bea
\left\langle \frac{\d \sigma_{q\bar{q}}^A}{\d^2 {\bm b}} \right\rangle &=&
2 \int \prod_{i=1}^A \d^2 {\bm b}_i T_A({\bm b}_i)  
\left[ 1 - \exp \left( -r^2 F(x,r) \sum_{i=1}^A T_p({\bm b} - {\bm b}_i) \right) \right]
=
2 [ 1 - (1-I({\bm b}))^A ].
\label{lumpy}
\eea
by factoring the $A$ integrals, where
\bean
I({\bm b}) &=& \int \d^2 {\bm b}^\prime T_A({\bm b} + {\bm b}^\prime )
\left[ 1 - \exp \left( -r^2 F(x,r)  T_p({\bm b}^\prime ) \right) \right].
\eean
Observing that the integral over ${\bm b}^\prime$ gains most contribution
over the size of the proton, which is small compared to the nucleus. Pursuing
this, we suppose $T_A({\bm b} + {\bm b}^\prime ) \approx T_A({\bm b})$. 
\bea
I ( {\bm b}) &\approx&
T_A ( {\bm b} ) 2\pi B_p \left[
\gamma - 
\textrm{Ei}\left( - r^2 F(x,r) / 2 \pi B_p \right) + 
\ln \left( r^2 F(x,r)/ 2 \pi B_p \right) \right].
\label{ln1}
\eea
Here $\gamma$ is the Euler-Mascheroni constant and Ei is the exponential
integral.
Fig.~\ref{fig2} shows the differential cross sections produced using these various
models. The size of the nucleus grows (in apparent physical size) as $x$ shrinks,
visible from the dip and peak positions. Within the GM models, a more sophisticated
lumpy calculation is almost indistinguishable from the smooth approximation. The
$\theta$-CK models overestimates the cross section, as expected. The GM models
reproduce an, on average, lower normalisations than rcBK.

\section{Conclusion}

Fig.~\ref{fig2} is our main result. The two
approaches, namely DGLAP and rcBK evolution, were checked against existing data for
electron-hadron collisions. They both gave reasonable agreement in the relevant $x$
range. However, when tested with electron-ion collisions the predictions appear
different, both in normalisation and peak-position. 

This work is ongoing, and many improvements are to be made. In particular,  
it is a priority to include ``nuclear shadowing''
effects by making use of newer nPDF software. Another avenue to explore would be to
repeat the above for $\phi$-meson scattering, as the larger dipole size $r$ makes for
a more sensitive probe of the saturation region.

\section{Acknowledgments}

G.S.J. wishes to thank the sponsors of the Hard Probes conference for entirely
funding his attendance. W.A.H. gratefully acknowledges the support of the
SA-CERN collaboration and the SA National Research Foundation.



%% References


\bibliographystyle{elsarticle-num}
\bibliography{bibliography.bib}






% \end{thebibliography}

\end{document}